# A Simplification Method of Polymorphic Boolean Functions

Wenjian Luo and Zhifang Li

*Abstract*—Polymorphic circuits are a special kind of circuits which possess multiple build-in functions, and these functions are activated by environment parameters, like temperature, light and VDD. The behavior of a polymorphic circuit can be described by a polymorphic Boolean function. For the first time, this brief presents a simplification method of the polymorphic Boolean function.

## I. INTRODUCTION

Compared with the traditional electronics, the polymorphic electronic components possess multiple build-in functions which are activated by environmental signals. For instance, the AND/OR polymorphic logic gate controlled by temperature perform the AND function when the temperature is 27°C and perform the OR function when the temperature is 125°C [1]. Due to the multiple-functional and sensing properties, polymorphic electronics have several potential applications in security, verification, multi-functional circuits and smart systems [2-4]. The possible applications of polymorphic electronics have been summarized in [1, 4]

Polymorphic gates are fundamental components of Polymorphic Electronics. They are basic building blocks of polymorphic circuits. In each working mode, a polymorphic gate would perform a different logic function. Some polymorphic gates have been designed and fabricated. In [5] and [6], two different NAND/NOR gate controlled by voltage are designed, and they are fabricated in silicon with 0.5 μm and 0.7 μm CMOS technology, respectively. Different AND/OR gates, which are controlled by temperature and voltage, respectively, are designed in [1] and [7]. NAND/NOR/NXOR/AND is reported in [8]. A reconfigurable polymorphic chip REPOMO32 is introduced in [9]. This REPOMO32 chip consists of 32 two-input elements which would performs the function AND, OR, XOR or NAND/NOR.

Polymorphic logic circuits are composed of polymorphic logic gates. In each working mode, a polymorphic circuit performs a traditional Boolean function. For example, a polymorphic circuit "4-parity / 4-majority" would perform as 4-bit parity in the first mode, and perform as 4-majority in the second mode.

Researchers have proposed some methods for synthesizing polymorphic circuits. Evolutionary methods are widely adopted [4, 10, 11]. The evolutionary methods could design area-efficient polymorphic circuits, but the Evolutionary Algorithms face the scalability problem. Therefore, only small scale circuits can be generated. Up to now, "3×4 multiplier / 7 bit sorting-net" reported in [4] is the biggest polymorphic circuits designed by evolutionary methods. In [12, 13], Sekanina and his colleagues proposed the Poly-BDD and polymorphic multiplex methods for designing polymorphic circuits. It is noted that polymorphic multiplex method is firstly proposed in [11], and also adopted in [14], but they are somewhat different. the In [15], the Poly_Bi_Decomposition method is proposed to synthesize polymorphic circuits.

The behavior of a polymorphic circuit can be described by a polymorphic Boolean function. The simplification of a Boolean function is common, and its importance stems from the fact that "the simpler the function is, the easier it is to realize" [16]. In this brief, a simplification method of polymorphic Boolean functions is given. It is shown that how the proposed simplification rules are applied to minimize a polymorphic Boolean function through Karnaugh Map [16, 17].

The rest of the paper is organized as follows. Section II introduces the simplification method of polymorphic Boolean functions. Section III shows examples of the simplification, and some guidelines are given for simplifying polymorphic Boolean function on Karnaugh Map. Section IV gives some discussion. Section V concludes this brief.

## II. SIMPLIFICATION OF POLYMORPHIC BOOLEAN FUNCTIONS

In this section, firstly, the definitions of the polymorphic Karnaugh Map and the polymorphic cube are introduced. Secondly, the simplification rules of polymorphic Boolean functions are given, and it is shown that how these rules are applied in the polymorphic Karnaugh Map.

### A. The Polymorphic Karnaugh Map and Polymorphic Cube

A polymorphic Boolean function $f$ can be presented as $f_1/f_2$, where $f_1$ and $f_2$ are traditional Boolean functions. In mode 1, its function is $f_1$, and in mode 2, its function is $f_2$. Figure 1 shows the polymorphic Boolean function of "4-parity / 4-majority". In mode 1, the function is 4 bit parity, and in mode 2, the function is 4 bit majority.

| $x_1 x_2$ \ $x_3 x_4$ | 00 | 01 | 11 | 10 |
|---|---|---|---|---|
| 00 | 0 | 1 | 0 | 1 |
| 01 | 1 | 0 | 1 | 0 |
| 11 | 0 | 1 | 0 | 1 |
| 10 | 1 | 0 | 1 | 0 |

4 bit parity

| $x_1 x_2$ \ $x_3 x_4$ | 00 | 01 | 11 | 10 |
|---|---|---|---|---|
| 00 | 0 | 0 | 0 | 0 |
| 01 | 0 | 0 | 1 | 0 |
| 11 | 0 | 1 | 1 | 1 |
| 10 | 0 | 0 | 1 | 0 |

4 bit majority

Wenjian Luo and Zhifang Li are with the Anhui Key Laboratory of Software in Computing and Communication, the School of Computer Science and Technology, University of Science and Technology of China, Hefei 230027, China (phone:86-551-3602824). Email: wjluo@ustc.edu.cn, zhifangl@mail.ustc.edu.cn.

Fig. 1. The polymorphic Boolean function "4-parity / 4-majority"

A polymorphic Boolean function can be expressed in the form of an Algebraic Expression. For example, $f = ((\bar{x}_2 \text{ AND/AND } x_3) \text{ XOR/OR } x_1) \text{ OR/OR } \bar{x}_4$ is the algebraic expression of a polymorphic Boolean function. For convenience, it can be rewritten as $f = \bar{x}_2 x_3 \text{ XOR/OR } x_1 + \bar{x}_4$. In the first mode, the function is $((\bar{x}_2 x_3) \oplus x_1) + \bar{x}_4$, and in the second mode, the function is $\bar{x}_2 x_3 + x_1 + \bar{x}_4$

Similar to the Karnaugh Map of traditional Boolean function, a polymorphic Boolean function can be described by a polymorphic Karnaugh Map. Figure 2 illustrates the polymorphic Karnaugh Map of "4-parity / 4-majority". Each square has four possible values, i.e. 0/0, 0/1, 1/0 and 1/1.

| $x_1 x_2$ \ $x_3 x_4$ | 00 | 01 | 11 | 10 |
|---|---|---|---|---|
| 00 | 0/0 | 1/0 | 0/0 | 1/0 |
| 01 | 1/0 | 0/0 | 1/1 | 0/0 |
| 11 | 0/0 | 1/1 | 0/1 | 1/1 |
| 10 | 1/0 | 0/0 | 1/1 | 0/0 |

Fig. 2. The polymorphic Karnaugh Map of "4-parity / 4-majority"

$B^n$ ($B \in \{0, 1\}$) is a hypercube described by the Boolean variables $x_1, x_2, \ldots, x_n$. If each node of a hypercube is assigned a binary value (i.e. 0 or 1), this hypercube specifies a Boolean function. Similarly, if each node is assigned a polymorphic value (i.e. 0/0, 0/1, 1/0 or 1/1), this hypercube specifies a polymorphic Boolean function. Figure 3 is the cube presentation of "4-parity / 4-majority".

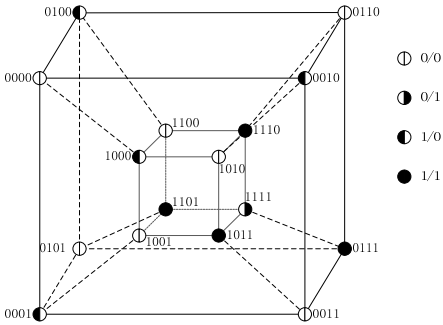

Fig. 3. The cube presentation of "4-parity / 4-majority"

A $m$-dimensional subcube is of paramount importance in the Boolean algebra. In fact, the characteristic function of a subcube (i.e. the points in the subcube are assigned value 1, and value 0 elsewhere) can be expressed as the product of $n - m$ variables. In the next section, based on the polymorphic cubes, the simplification rules of polymorphic Boolean function are given.

*B. The Simplification Rules*

Figure 4 gives the basic rules to minimize a polymorphic Boolean function. In Figure 4, $c_1 = c_{1,1}/c_{1,2}$ and $c_2 = c_{2,1}/c_{2,2}$ are two polymorphic subcubes, where $c_{1,1}$ and $c_{2,1}$ are the cubes in the first mode, and $c_{2,1}$ and $c_{2,2}$ are the cubes in the second mode. The points in dashed part of the cubes have value 1, and the points in the white part of the cubes have value 0. $P_1$ and $P_2$ are the expression of characteristic function of $c_1$ and $c_2$, respectively.

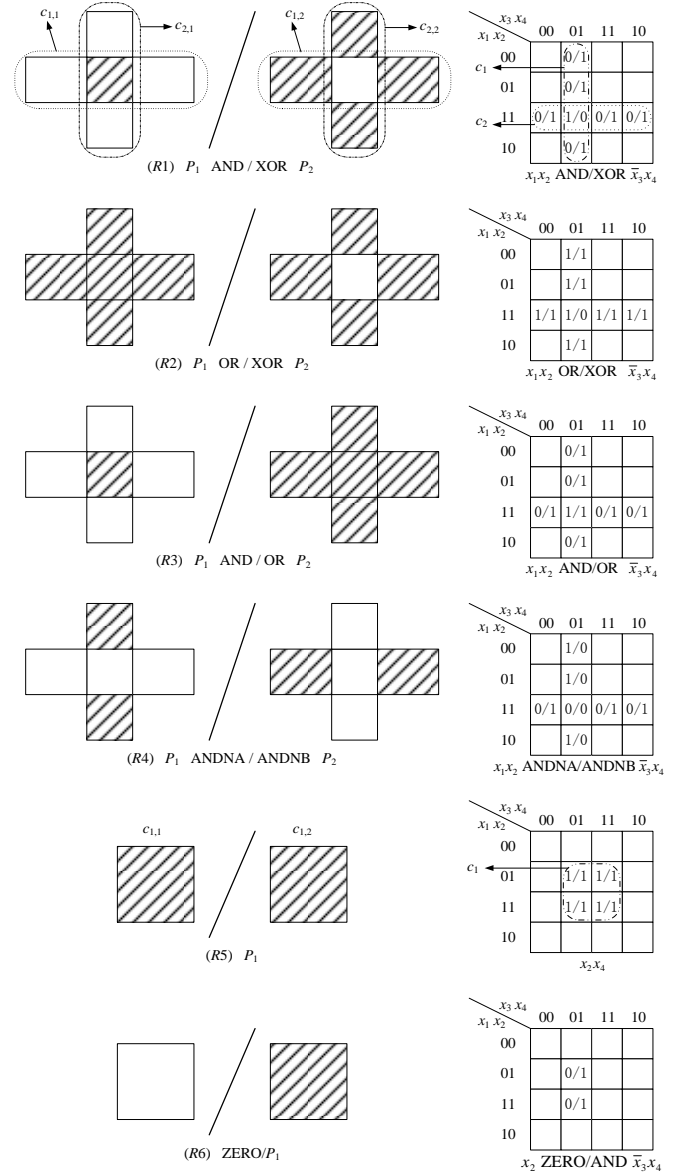

Fig. 4. The basic simplification rules of polymorphic Boolean functions

In Figure 4(*R*1), if $c_{1,1} \cap c_{2,1}$ is 1-cube (i.e. all points in the cube have value 1, $(c_{1,1} \cap c_{2,1}) - (c_{1,1} \cap c_{2,1})$ is 0-cube, $c_{1,2} \cap c_{2,2}$ is 0-cube and $(c_{1,2} \cap c_{2,2}) - (c_{1,2} \cap c_{2,2})$ is 1-cube, $c_1$ and $c_2$ can be expressed as "$P_1$ AND/XOR $P_2$". In the right of Figure 4, examples are given to show how these rules are applied in polymorphic Karnaugh Map.

In many cases, the minimization rules in Figure 4 are not sufficient. For example, the instance in Figure 5(*a*) is a modified version of the pattern in Figure 4($R$1). Patterns in Figure 5($b$) and Figure 5($c$) are often met in the simplification. However, they cannot be minimized.

Therefore, the extended rules are given in Figure 6, where three polymorphic cubes are considered in the simplification.

Figure 6($R$7, $R$8), Figure 6($R$9, $R$10), Figure 6($R$11, $R$12) and Figure 6($R$13, $R$14) are the extended versions of Figure 4($R$1), Figure 4($R$2), Figure 4($R$3) and Figure 4($R$4), respectively. Figure 7($R$15) and Figure 7($R$16) solve the minimization of Figure 5($b$) and Figure 5($c$), respectively.

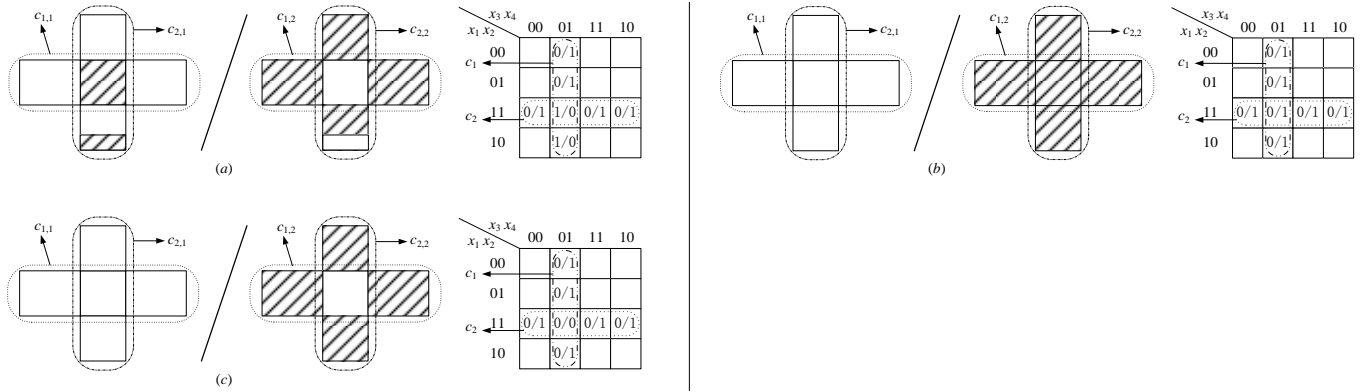

Fig. 5. The patterns that can not be simplified by $R$1 ~ $R$6

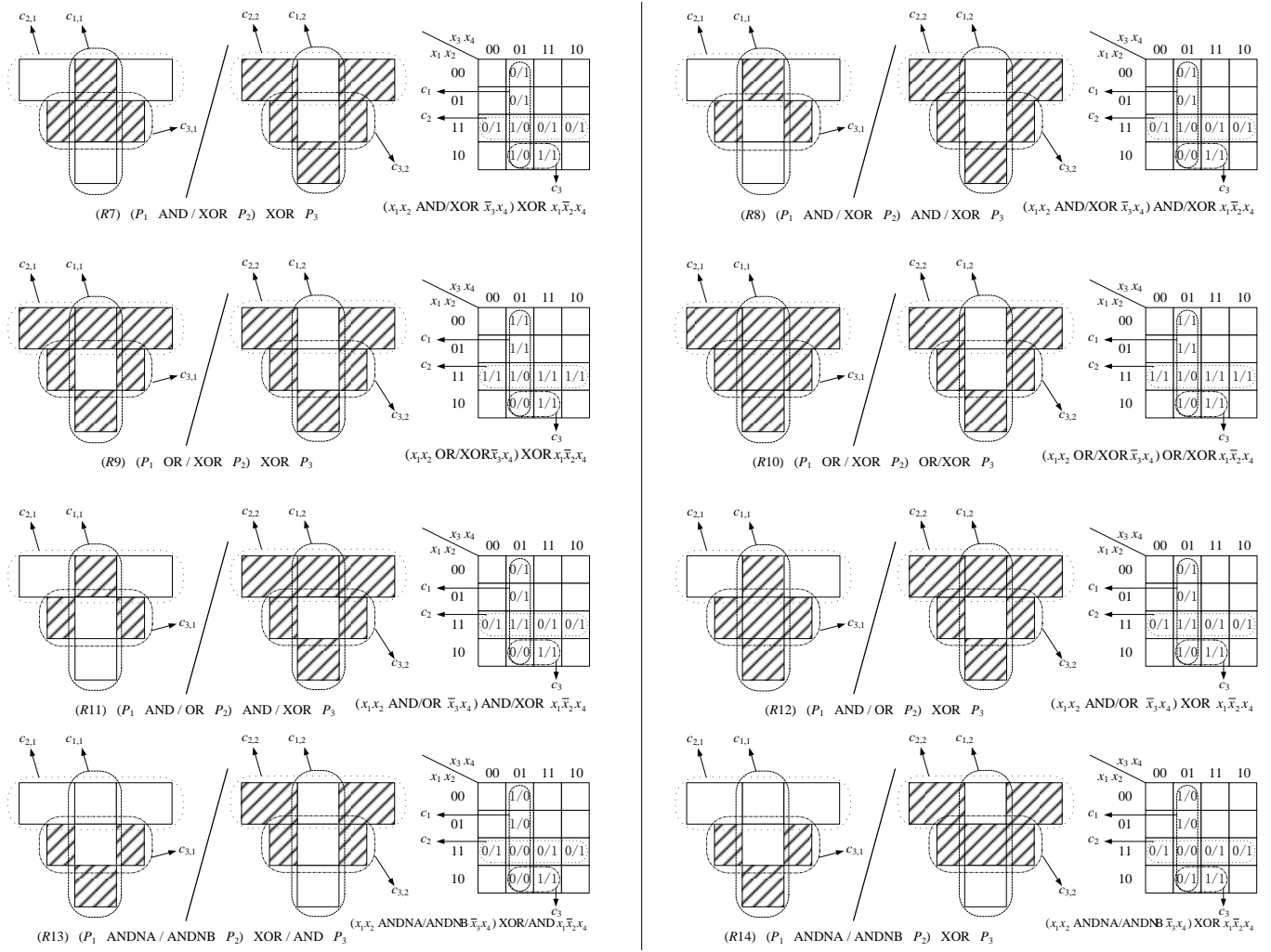

Fig. 6. The extended simplification rules of polymorphic Boolean functions

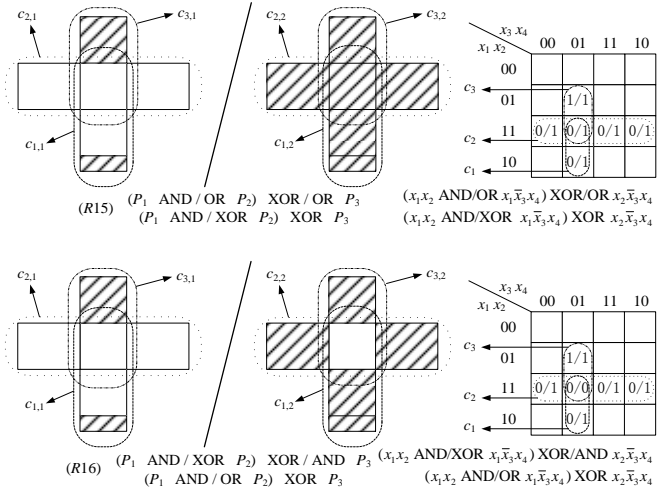

Fig. 7. The extended simplification rules of polymorphic Boolean functions

## III. SOME SIMPLIFICATION EXAMPLES

In this section, based on the polymorphic Karnaugh Map, the simplification rules introduced in Section II are applied to minimize some polymorphic Boolean functions.

Figure 8 shows the minimization of "4-parity / 4-majority". In the simplification, rule $R2$ in Figure 4 and rule $R16$ in Figure 6 are applied.

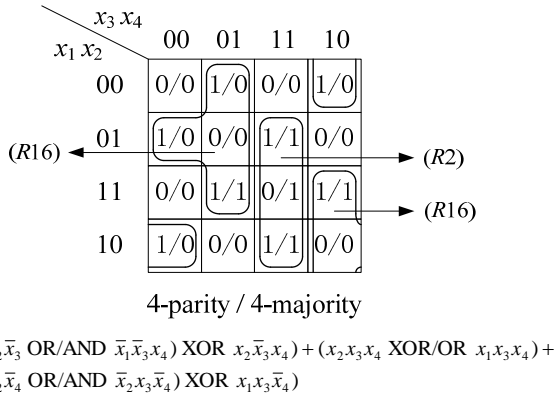

4-parity / 4-majority

$((\bar{x}_1 x_2 \bar{x}_3 \text{ OR/AND } \bar{x}_1 \bar{x}_3 x_4) \text{ XOR } x_2 \bar{x}_3 x_4) + (x_2 x_3 x_4 \text{ XOR/OR } x_1 x_3 x_4) +$
$((x_1 \bar{x}_2 \bar{x}_4 \text{ OR/AND } \bar{x}_2 x_3 \bar{x}_4) \text{ XOR } x_1 x_3 \bar{x}_4)$

Fig. 8. The simplification of the polymorphic Boolean function "4-parity / 4-majority"

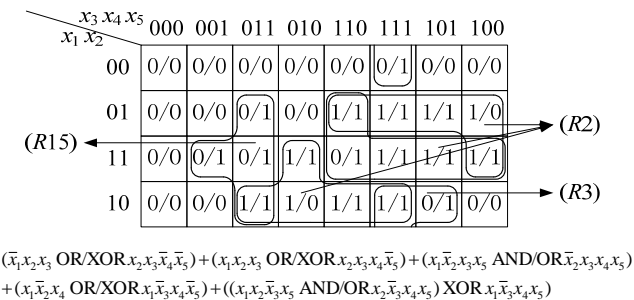

$(\bar{x}_1 x_2 x_3 \text{ OR/XOR } x_2 x_3 \bar{x}_4 \bar{x}_5) + (x_1 x_2 x_3 \text{ OR/XOR } x_2 x_3 x_4 \bar{x}_5) + (x_1 \bar{x}_2 x_3 x_5 \text{ AND/OR} \bar{x}_2 x_3 x_4 x_5)$
$+ (x_1 \bar{x}_2 x_4 \text{ OR/XOR } x_1 \bar{x}_3 x_4 \bar{x}_5) + ((x_1 x_2 \bar{x}_3 x_5 \text{ AND/OR} x_2 x_3 x_4 x_5) \text{ XOR } x_1 x_3 x_4 x_5)$

Fig. 9. The simplification of the polymorphic Boolean function "2×3 multiplier (3) / 5-sortingnet (3)"

Figure 9 shows the minimization of "2×3 multiplier (3) / 5-sortingnet (3)", where "2×3 multiplier (3)" means the third output of 2×3 multiplier, and "5-sortingnet (3)" means the third output of 5 bit sorting network. In the course of simplification, rule $R2$ and rule $R3$ in Figure 4 and rule $R15$ in Figure 6 are applied.

Karnaugh Map is a convenient and widely used tool for minimizing small scale Boolean functions (3 to 6 Boolean variables) manually. Due to the complex of the polymorphic Boolean function simplification rules, the simplification of polymorphic Boolean functions is more complicated than the simplification in the traditional Karnaugh Map. In [16], some guidance are given for simplification in traditional Karnaugh Map. Based on those guidelines, some rules are given below which should be kept in mind when simplifying polymorphic Boolean functions with the polymorphic Karnaugh Map.

1. Makes as few as groups to cover all the squares [16].
2. Groups as many squares together as possible [16].
3. Try to find the patterns listed in Figure 4. If two polymorphic cubes form a very similar but different pattern from the patterns in Figure 4, or two polymorphic cubes form the pattern in Figure 5(*b*) and Figure 5(*c*), try the extended patterns listed in Figure 6 and Figure 7.

## IV. DISCUSSION

The simplification of polymorphic Boolean functions is complicated compared with the traditional single mode Boolean function. As for the minimization rules given in Figure 4, Figure 6 and Figure 7, two or three polymorphic cubes are considered simultaneously.

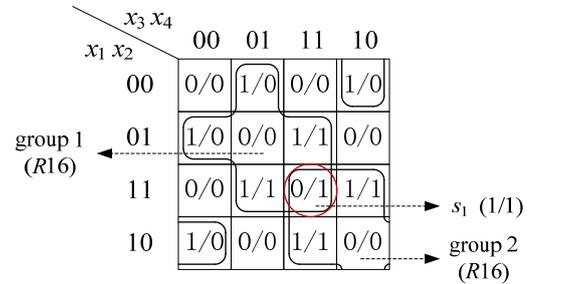

$((x_1 \bar{x}_2 \bar{x}_4 \text{ OR/AND} \bar{x}_2 x_3 \bar{x}_4) \text{ XOR } x_1 x_3) \text{ XOR/OR} ((\bar{x}_1 x_2 \bar{x}_3 \text{ OR/AND} \bar{x}_1 \bar{x}_3 x_4) \text{ XOR } x_2 x_4)$

Fig. 9. The simplification of the polymorphic Boolean function "4-parity / 4-majority"

In fact, if more cubes (>3) are considered, and those simplification rules (in Figure 4, 6 and 7) are applied simultaneously, more compact expression can be obtained.

For example, as for the "4-parity / 4-majority" (depicted in Figure 9), if the square "$s_1$" is set to "1/1" temporarily, the simplification rule $R16$ in Figure 7 can be adopted in "group 1" and "group 2". Then, the value of "$s_1$" square is set back to "0/1", and the simplification rule $R2$ in Figure 4 can be adopted on "group 1" and "group 2".

However, it is very hard to operate the simplification on many cubes simultaneously. In the future, the algorithms for the simplification of polymorphic Boolean function will be studied.

Traditional Boolean function minimization tools, such as Espresso [18], express the Boolean function in the sum of product (SOP) form, which can be used to synthesize two-level or multi-level circuits. In this brief, due to the intrinsic complexity of polymorphic Boolean function, many kinds of polymorphic gates are adopted in the simplification. Therefore, after the simplification with the proposed rules, the expression can be used to synthesize multi-level polymorphic circuits.

## V. CONCLUSION

In this brief, some simplification rules for minimizing polymorphic Boolean function are given. It is shown that how these rules are applied on polymorphic Karnaugh Map. Compared with the simplification rules of traditional Boolean function for synthesizing two-level circuits, the rules proposed in this brief are used for the polymorphic Boolean functions. In the future, based on the rules given here, algorithms for minimizing polymorphic Boolean function will be studied.

ACKNOWLEDGMENT

This work is partly supported by the Fundamental Research Funds for the Central Universities.